\documentclass[prc,aps,superscriptaddress,floatfix,twocolumn,nofootinbib]{revtex4-2}

\usepackage{amsmath}
\usepackage{amssymb}
\usepackage{bm}
\usepackage{graphicx}
\usepackage{multirow}

\begin{document}

\title{\textit{Ab initio} no-core shell model calculations of the electric dipole polarizabilities and photoabsorption cross sections of the $A=3$ mirror nuclei}

\author{H. T. Zhao}
\affiliation{College of Physics and Engineering, Henan University of Science and Technology, Luoyang 471023, China}
\author{C. Y. Zhai}
\affiliation{College of Physics and Engineering, Henan University of Science and Technology, Luoyang 471023, China}
\author{J. P. Vary}
\affiliation{Department of Physics and Astronomy, Iowa State University, Ames, IA 50011, USA}
\author{H. Li}
\affiliation{CAS Key Laboratory of High Precision Nuclear Spectroscopy, Institute of Modern Physics, Chinese Academy of Sciences, Lanzhou 730000, China}
\affiliation{School of Nuclear Science and Technology, University of Chinese Academy of Sciences, Beijing 100049, China}
\author{J. M. Dong}
\affiliation{CAS Key Laboratory of High Precision Nuclear Spectroscopy, Institute of Modern Physics, Chinese Academy of Sciences, Lanzhou 730000, China}
\affiliation{School of Nuclear Science and Technology, University of Chinese Academy of Sciences, Beijing 100049, China}
\author{H. J. Ong}
\affiliation{CAS Key Laboratory of High Precision Nuclear Spectroscopy, Institute of Modern Physics, Chinese Academy of Sciences, Lanzhou 730000, China}
\affiliation{School of Nuclear Science and Technology, University of Chinese Academy of Sciences, Beijing 100049, China}
\author{X. Zhao}
\affiliation{CAS Key Laboratory of High Precision Nuclear Spectroscopy, Institute of Modern Physics, Chinese Academy of Sciences, Lanzhou 730000, China}
\affiliation{School of Nuclear Science and Technology, University of Chinese Academy of Sciences, Beijing 100049, China}

\author{\protect\\P. J. Fasano}
\affiliation{Department of Physics and Astronomy, University of Notre Dame, Notre Dame, IN 46556-5670, USA}
\author{A. M. Shirokov}
\affiliation{Skobeltsyn Institute of Nuclear Physics, Lomonosov Moscow State University, Moscow 119991, Russia}
\author{J. Chen}
\affiliation{Department of Physics, Southern University of Science and Technology, Shenzhen 518055, Guangdong, China}
\author{D. Y. Tao}
\affiliation{Key Laboratory of Nuclear Physics and Ion-Beam Application (MoE), Institute of Modern Physics, Fudan University, Shanghai 200433, China}
\affiliation{Shanghai Research Center for Theoretical Nuclear Physics, NSFC and Fudan University, Shanghai 200438, China}
\author{B. Zhou}
\affiliation{Key Laboratory of Nuclear Physics and Ion-Beam Application (MoE), Institute of Modern Physics, Fudan University, Shanghai 200433, China}
\affiliation{Shanghai Research Center for Theoretical Nuclear Physics, NSFC and Fudan University, Shanghai 200438, China}
\author{C. Ji}
\affiliation{Key Laboratory of Quark and Lepton Physics, Institute of Particle Physics, Central China Normal University, Wuhan 430079, China}
\affiliation{Southern Center for Nuclear-Science Theory (SCNT), Institute of Modern Physics, Chinese Academy of Sciences, Huizhou 516000, Guangdong Province, China}
\author{P. Yin}
\email{Corresponding author: pengyin@iastate.edu}
\affiliation{College of Physics and Engineering, Henan University of Science and Technology, Luoyang 471023, China}
\affiliation{State Key Laboratory of Heavy Ion Science and Technology, Institute of Modern Physics, Chinese Academy of Sciences, Lanzhou 730000, China}

\begin{abstract}
We present \textit{ab initio} no-core shell-model calculations of the electric dipole ($E1$) polarizabilities and photoabsorption cross sections of the $A=3$ mirror nuclei $^3$H and $^3$He using the Daejeon16 nucleon-nucleon interaction. The polarizabilities are obtained by direct summation over discretized continuum states in the harmonic-oscillator basis, while the same spectral information is used to reconstruct the cross sections. This work provides the first extension of the approach from the spin-zero $^4$He to nonzero-spin nuclei by calculating the $1/2^-$ and $3/2^-$ final-state contributions separately and demonstrating their model-space convergence. In both nuclei, the $3/2^-$ contribution is approximately twice the $1/2^-$ contribution. After quantifying finite-basis and excitation-energy-truncation uncertainties, we identify a clear difference between the mirror polarizabilities, induced entirely by the Coulomb interaction because the Daejeon16 interaction is isospin symmetric. The reconstructed cross sections remain stable under variations of the inversion parameters of the Lorentz integral transform and exhibit the same angular-momentum hierarchy. These results extend the discretized-continuum approach to integrated and energy-resolved $E1$ observables in nonzero-spin nuclei with multiple final-state angular momenta.
\end{abstract}

\maketitle

\section{Introduction}

Electromagnetic response functions provide direct access to the dynamical structure of atomic nuclei.  Of particular interest is the electric dipole ($E1$) polarizability $\alpha_E$, the inverse-energy-weighted sum rule of the continuum $E1$ response.  The enhanced weight assigned to low excitation energies makes $\alpha_E$ sensitive to breakup thresholds, long-range correlations, and the spatial distribution of charge.  The underlying $E1$ strength also determines the $E1$ photoabsorption cross section, so that the polarizability and the energy-dependent cross section probe complementary aspects of the same nuclear response.

The principal theoretical challenge is that the $E1$ response extends into the nuclear continuum.  Within \textit{ab initio} nuclear theory, several complementary strategies have been developed to account for this continuum strength.  Exact three-body continuum approaches, including Faddeev and Alt-Grassberger-Sandhas (AGS) formulations, solve the scattering and breakup problems explicitly \cite{Barbour1970,Gibson1975,Fang1978,Vostrikov1981,Golak2000,Skibinski2003,Skibinski2005,Deltuva2004,Deltuva2005}.  Stieltjes and Lorentz integral transforms map the continuum response onto localized many-body equations, thereby avoiding the explicit construction of scattering wave functions \cite{Miorelli2016,Efros1994,Efros1997,EfrosReview2007}.  Inhomogeneous-equation and Lanczos sum-rule techniques provide an efficient way to calculate response moments without explicitly determining every continuum state. These approaches have been employed in no-core shell model (NCSM) calculations~\cite{Stetcu2009}, effective-interaction hyperspherical harmonics studies~\cite{NevoDinur2016,NevoDinur2014}, and symmetry-adapted NCSM calculations~\cite{Baker2020}.  Experimentally, the polarizability can be obtained by integrating the measured photoabsorption cross section with the nverse-energy-squared weight \cite{Rinker1976,PachuckiMoro2007}. It can also be extracted from nuclear scattering measurements at energies below the Coulomb barrier~\cite{Goeckner1991}.

An alternative direct approach was suggested in Ref.~\cite{Yin2024}.  In that work, a large number of discretized continuum states were obtained by diagonalizing the Hamiltonian in the NCSM harmonic-oscillator (HO) basis, and the $^4$He $E1$ polarizability was calculated directly from their excitation energies and $E1$ transition strengths.  The continuum is discretized within the finite model space, and the reliability
of this approximation is supported by the convergence of the calculated polarizability with increasing model-space size.  Since the individual excitation energies and transition strengths are retained, the contributions from different energy regions and their convergence with model-space size can be examined directly.  More recently, the same discrete spectral information was used to construct and invert the Lorentz integral transform (LIT) and thereby obtain the $^4$He photoabsorption cross section \cite{Yin2026}.  These two developments provide, respectively, a direct calculation of an integrated response and an energy-resolved reconstruction based on the same many-body eigenstates.

Previous applications of this approach were restricted to $^4$He.  Since its ground state has $J_0^\pi=0^+$, the $E1$ transitions populate only $J_f^\pi=1^-$ final states.  Consequently, the reliability of this implementation has not yet been established for a nucleus in which the $E1$ response is distributed over several final-state angular momenta that should be calculated independently.  The $A=3$ nuclei provide a natural testing ground for this extension: their $J_0^\pi=1/2^+$ ground states couple through the $E1$ operator to both $J_f^\pi=1/2^-$ and $3/2^-$ final states.  The two sets of calculated states have different dimensions and level densities in a truncated basis.  Demonstrating that the $E1$ polarizability contribution from each sector stabilizes with increasing model-space size is therefore essential for validating the direct calculation when multiple final-state angular momenta contribute.  The decomposition also reveals how the $E1$ polarizability is distributed between the two allowed final-state angular momenta.

The mirror relation between $^3$H and $^3$He supports a further physics motivation.  In the charge-symmetry limit, interchange of protons and neutrons changes the sign of the intrinsic $E1$ operator but leaves the corresponding squared matrix elements unchanged; the two nuclei would consequently have identical $E1$ polarizabilities \cite{Stetcu2009,Miller1990,Miller2006}.  Their difference therefore provides a particularly direct measure of isospin-breaking effects in the low-energy $E1$ response.  Owing to the inverse-energy weighting, the polarizability is especially sensitive to symmetry-breaking shifts of strength near the breakup threshold.  Previous theoretical calculations of the $A=3$ $E1$ polarizabilities have employed hyperspherical-harmonics methods~\cite{Efros1997,NevoDinur2016} and the NCSM~\cite{Stetcu2009}.

In this work, we calculate the $^{3}$H and $^{3}$He $E1$ polarizabilities by summing the contributions of explicitly obtained negative-parity states, establish the convergence of the $1/2^-$ and $3/2^-$ components independently, and quantify the numerical uncertainties associated with the basis-space truncation and with the upper energy limit in calculations of $E1$ excitation.  We then use the same excitation energies and transition strengths to reconstruct the corresponding $E1$ photoabsorption cross sections.  This unified treatment extends the approach approbated in $^4$He studies~\cite{Yin2024,Yin2026} to nonzero-spin systems with multiple final-state angular momenta and enables a consistent comparison of the integrated and energy-dependent $E1$ responses of a mirror pair.

\section{Theoretical framework}

We use the \textit{ab initio}  NCSM \cite{Barrett2013} to calculate the nuclear energy spectra and wave functions.  In the NCSM, all nucleons are active, and the chosen nuclear Hamiltonian is diagonalized in a truncated HO Slater-determinant basis.  We perform the calculations using the Daejeon16 nucleon--nucleon ($NN$) interaction \cite{Shirokov2016}, which provides a good description of many properties of light nuclei, and add the Coulomb interaction between protons.  The many-body diagonalizations and $E1$ matrix-element evaluations are carried out with the Many-Fermion Dynamics--nuclear (MFDn) code \cite{MarisMFDn2010,AktulgaMFDn2014,ShaoMFDn2018,MarisMFDnGPU2022,CookMFDn2022,FasanoMFDnTransitions2025}.

The diagonalizations are carried out in an $M$-scheme basis with fixed
angular-momentum projection
$M=\sum_i m_i$ and parity.  The HO energy is denoted by $\hbar\Omega$, and the basis size is specified by $N_{\max}$, the maximum number of HO excitation quanta above the lowest Pauli-allowed configuration.
The negative-parity final states are calculated in model spaces with odd values of
$N_{\max}$, $N_{\max}=1,3,5,\ldots$; for each such calculation, the positive-parity
ground state is obtained at $N_{\max}-1$. All results are labeled by the final-state
value of $N_{\max}$. To highlight the approach to model-space convergence, we present
only the results for $N_{\max}=3,5,\ldots,19$. Spurious center-of-mass excitations are
removed with the Lawson prescription \cite{GloecknerLawson1974,Whitehead1977}.

The $J_f^\pi=1/2^-$ and $3/2^-$ spectra are obtained in separate diagonalizations.  For a selected $J_f$, we work in the $M=J_f$ sector.  To minimize the use of computational resources, states with $J>J_f$ are shifted away by adding the angular-momentum filter $H_{J^2}=\lambda[\bm J^2-J_f(J_f+1)]$ during the diagonalization \cite{Yin2024,Whitehead1977}.  This term leaves the desired $J=J_f$ states unchanged.  The value of $\lambda$ is chosen so that the unwanted $J>J_f$ states lie above the adopted cutoff $E_x^{\rm cut}$, which defines the upper excitation energy retained in both the direct polarizability sum and the finite-spectrum LIT input.

For the $k$th calculated excited state, the reduced $E1$ strength is
\begin{equation}
 \begin{aligned}
 B_k(E1)
 &\equiv B(E1;J_0\!\to\!J_k)\\
 &=\sum_{M_k\mu}\left|
 \langle\Psi_k^{J_kM_k}|\hat D_{1\mu}|\Psi_0^{J_0M_0}\rangle
 \right|^2\\
 &=\frac{\left|\langle\Psi_k^{J_k}\|\hat D_1\|\Psi_0^{J_0}\rangle\right|^2}{2J_0+1}.
 \end{aligned}
 \label{eq:BE1}
\end{equation}
In the long-wavelength approximation, we employ the one-body $E1$ operator used in Ref.~\cite{Yin2026},
\begin{equation}
 \hat D_{1\mu}=\sum_{i=1}^{A}q_i r_iY_{1\mu}(\hat{\bm r}_i),
 \label{eq:D}
\end{equation}
where $q_i$ is the charge of nucleon $i$, and $\bm r_i$ is its single-particle coordinate, with $r_i=|\bm r_i|$ and $\hat{\bm r}_i=\bm r_i/r_i$.  The excitation energy of the $k$th state is $E_{x,k}=E_k-E_0$.  The $E1$ polarizability, evaluated with an excitation-energy cutoff $E_x^{\rm cut}$, is
\begin{equation}
 \alpha_E(E_x^{\rm cut})=\frac{8\pi}{9}\alpha_{\rm em}
 \sum_{0<E_{x,k}\le E_x^{\rm cut}}
 \frac{B_k(E1)}{E_{x,k}},
 \label{eq:alpha}
\end{equation}
where $\alpha_{\rm em}$ is the fine-structure constant.  The total polarizability includes all calculated $J_f^\pi=1/2^-$ and $3/2^-$ states in the sum.  We use $E_x^{\rm cut}=100$ MeV; the sufficiency of this truncation is examined in Sec.~III.  We use natural units, $\hbar=c=1$, and express the final polarizabilities and photoabsorption cross sections in fm$^3$ and mb, respectively.

The photoabsorption calculation follows the finite-spectrum LIT framework developed within the NCSM in Ref.~\cite{Yin2026}.  The inclusive $E1$ response is written formally as
\begin{equation}
 R(E_\gamma)=\sum_k B(E1;J_0\!\to\!J_k)\,
 \delta(E_\gamma-E_{x,k}),
 \label{eq:response}
\end{equation}
where $E_\gamma$ is the photon energy, and the sum runs over all allowed final states.  In the NCSM, the calculated states entering the LIT are retained below the same cutoff $E_x^{\rm cut}$.  The corresponding transform is
\begin{align}
 L(\sigma_R,\sigma_I)
 &=\int dE_\gamma\,
 \frac{R(E_\gamma)}
 {(E_\gamma-\sigma_R)^2+\sigma_I^2}
 \nonumber\\
 &\approx\sum_{E_{x,k}\le E_x^{\rm cut}}
 \frac{B_k(E1)}
 {(E_{x,k}-\sigma_R)^2+\sigma_I^2},
 \label{eq:LIT}
\end{align}
where $\sigma_R$ and $\sigma_I>0$ are the real and imaginary Lorentz parameters.  All allowed $E1$ transitions are included in constructing the LIT for the total response.  The photoabsorption cross section is obtained from the reconstructed response as
\begin{equation}
 \sigma_\gamma(E_\gamma)=\frac{16\pi^3}{9}\,
 \alpha_{\rm em}E_\gamma R(E_\gamma).
 \label{eq:sigma}
\end{equation}
The inversion procedure follows Ref.~\cite{Yin2026}; the specific inversion basis and numerical parameters used here are given in Appendix~\ref{app:LIT}.

\section{Results and discussion}

\begin{figure}[!tbp]
 \centering
 \includegraphics[width=1.0\columnwidth]{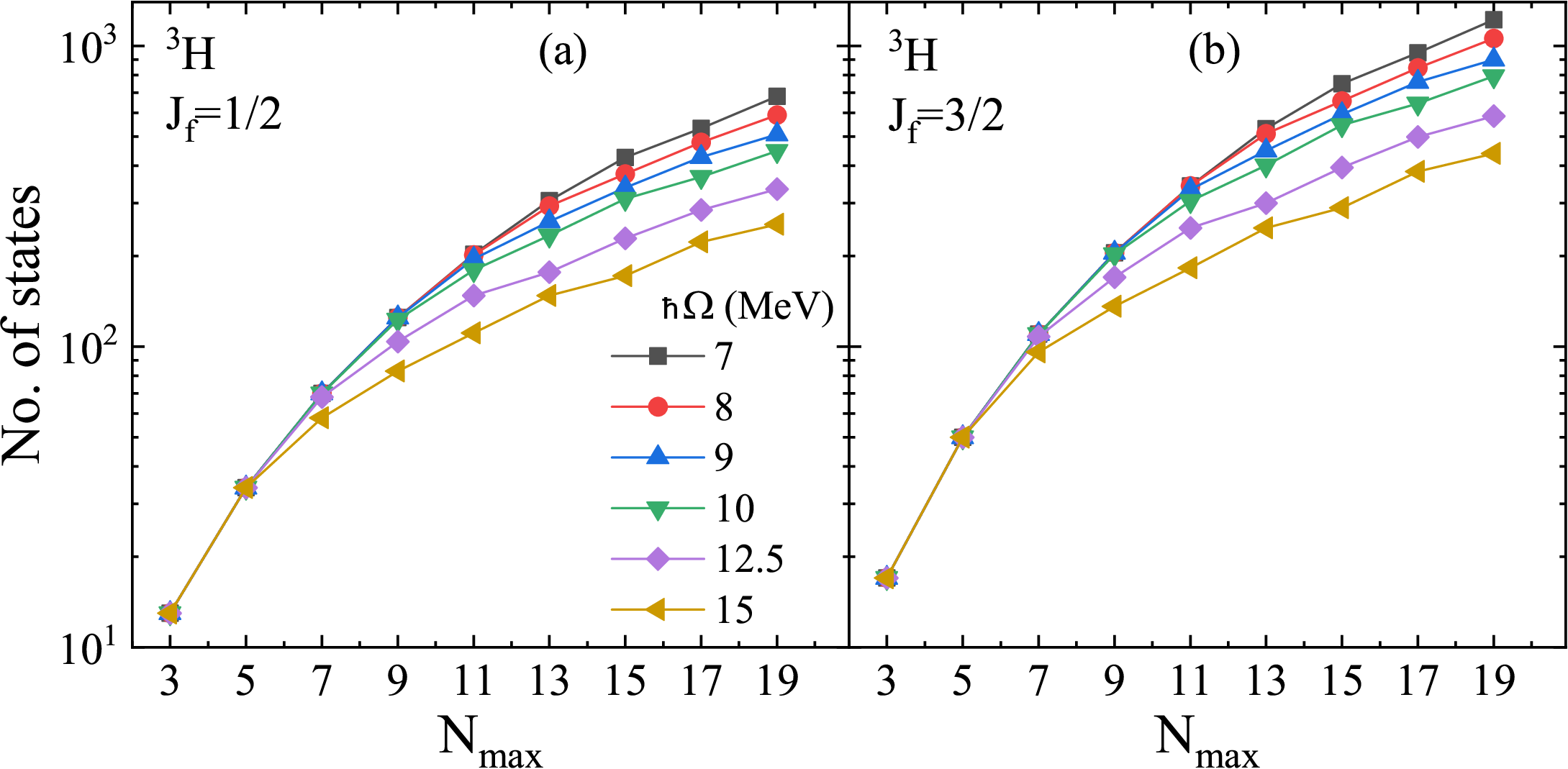}
 \caption{Numbers of negative-parity states below $E_x^{\rm cut}=100$ MeV as functions of the final-state $N_{\max}$ at the indicated HO energies.  Panel (a) corresponds to the $J_f^\pi=1/2^-$ sector of $^3$H, and panel (b) corresponds to the respective $J_f^\pi=3/2^-$ sector.}
 \label{fig:states}
\end{figure}

The number of retained negative-parity states of $^{3}\mathrm{H}$ is shown in Fig.~\ref{fig:states}.  For both spin-parities, $1/2^-$ and $3/2^-$, the number of states grows rapidly with increasing $N_{\max}$, and this growth is particularly pronounced at smaller $\hbar\Omega$.  This behavior closely parallels that observed in the earlier $^4$He calculations \cite{Yin2024,Yin2026}.  At fixed $N_{\max}$, the $3/2^-$ sector contains substantially more states than the $1/2^-$ sector. Because the ${}^{3}\mathrm{H}$ and ${}^{3}\mathrm{He}$ calculations employ mirror model spaces with identical dimensions, their state-count systematics are nearly identical. We therefore show only the ${}^{3}\mathrm{H}$ results in Fig.~1. The slight differences in the numbers of $^3$H and $^3$He states retained below $E_x^{\mathrm{cut}}=100~\mathrm{MeV}$ arise from Coulomb-induced shifts of the excitation energies.

\begin{figure}[!tbp]
 \centering
 \includegraphics[width=1.0\columnwidth]{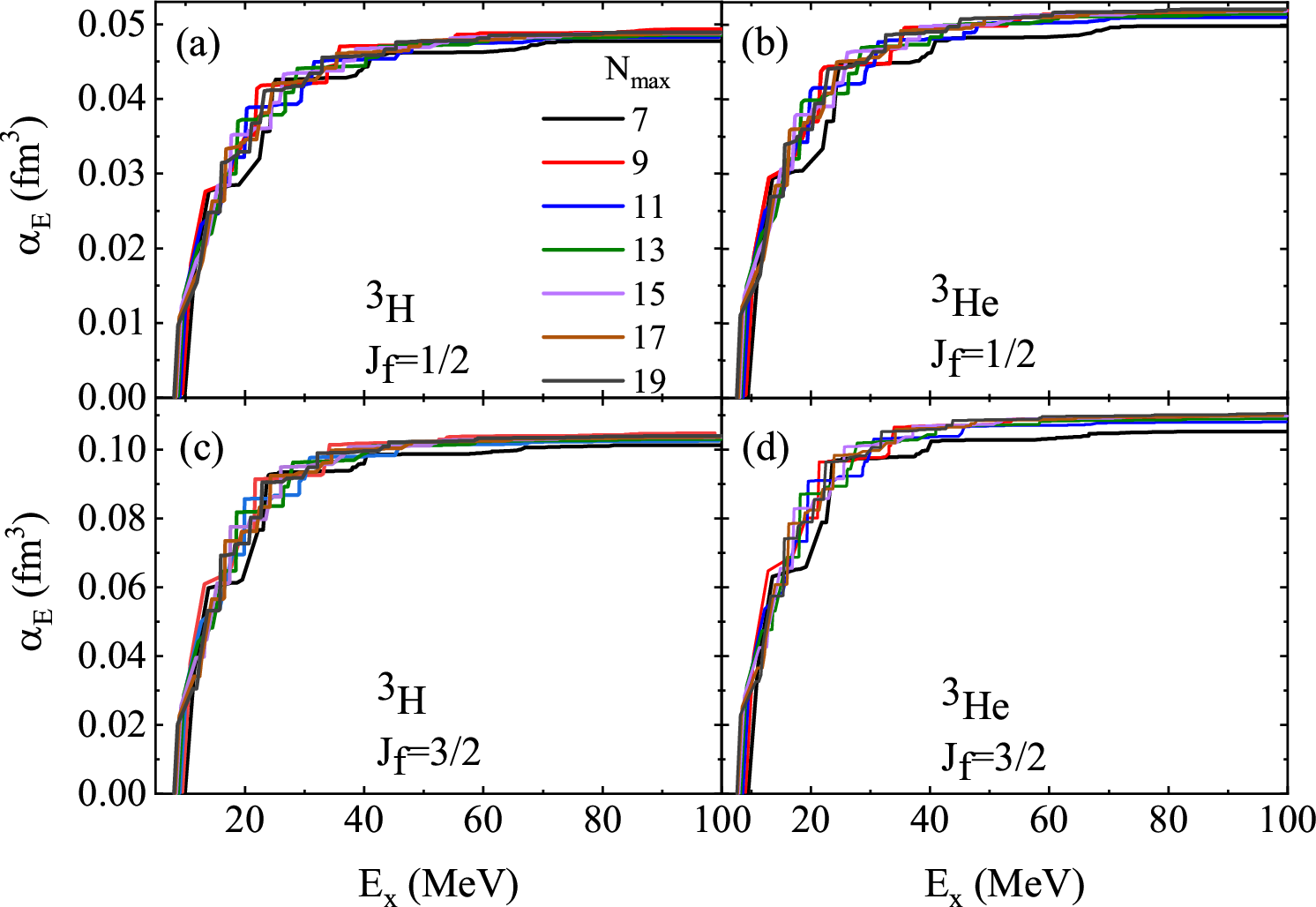}
 \caption{Channel-resolved running polarizabilities $\alpha_E(E_x)$ at $\hbar\Omega=10$ MeV for increasing $N_{\max}$.  Panels (a) and (b) show the $J_f^\pi=1/2^-$ contributions for $^3$H and $^3$He, and panels (c) and (d) show the corresponding $J_f^\pi=3/2^-$ contributions.}
 \label{fig:running}
\end{figure}

We trace the running polarizability in each angular-momentum sector in Fig.~\ref{fig:running}. Most of the inverse-energy-weighted strength is exhausted below 40 MeV.  Despite the markedly different state counts in Fig.~\ref{fig:states}, the endpoint contributions from the $J_f^\pi=1/2^-$ and $3/2^-$ sectors progressively stabilize as $N_{\max}$ increases.  Their ratio $\mathcal R_J\equiv\alpha_E^{3/2^-}/\alpha_E^{1/2^-}$ approaches a value near two; at $N_{\max}=19$ and $E_x^{\rm cut}=100$ MeV, it is about $2.12$ for both $^3$H and $^3$He.  Because the $E1$ operator is independent of spin variables, it conserves the intrinsic spin component in the transition. Consequently, for states belonging to the same orbital-spin coupling scheme, the summed $E1$ strength is governed by the magnetic-state multiplicity of the final state. Simple angular-momentum considerations therefore suggest a proportionality between the $E1$ strength (and the corresponding polarizability contribution) and the final-state degeneracy factor $2J_f+1$, yielding $\alpha_E^{3/2^-}:\alpha_E^{1/2^-}=2:1$.  Configuration mixing and state-dependent correlations make this only a qualitative reference, not an exact NCSM constraint.  The observed convergence and stable channel hierarchy demonstrate that the $3/2^-$ dominance is a robust feature of the present $A=3$ response; whether the near-$2:1$ ratio is a general consequence of quantum-mechanical final-state multiplicity would be interesting to test in other nuclei.

\begin{figure}[!tbp]
 \centering
 \includegraphics[width=1.0\columnwidth]{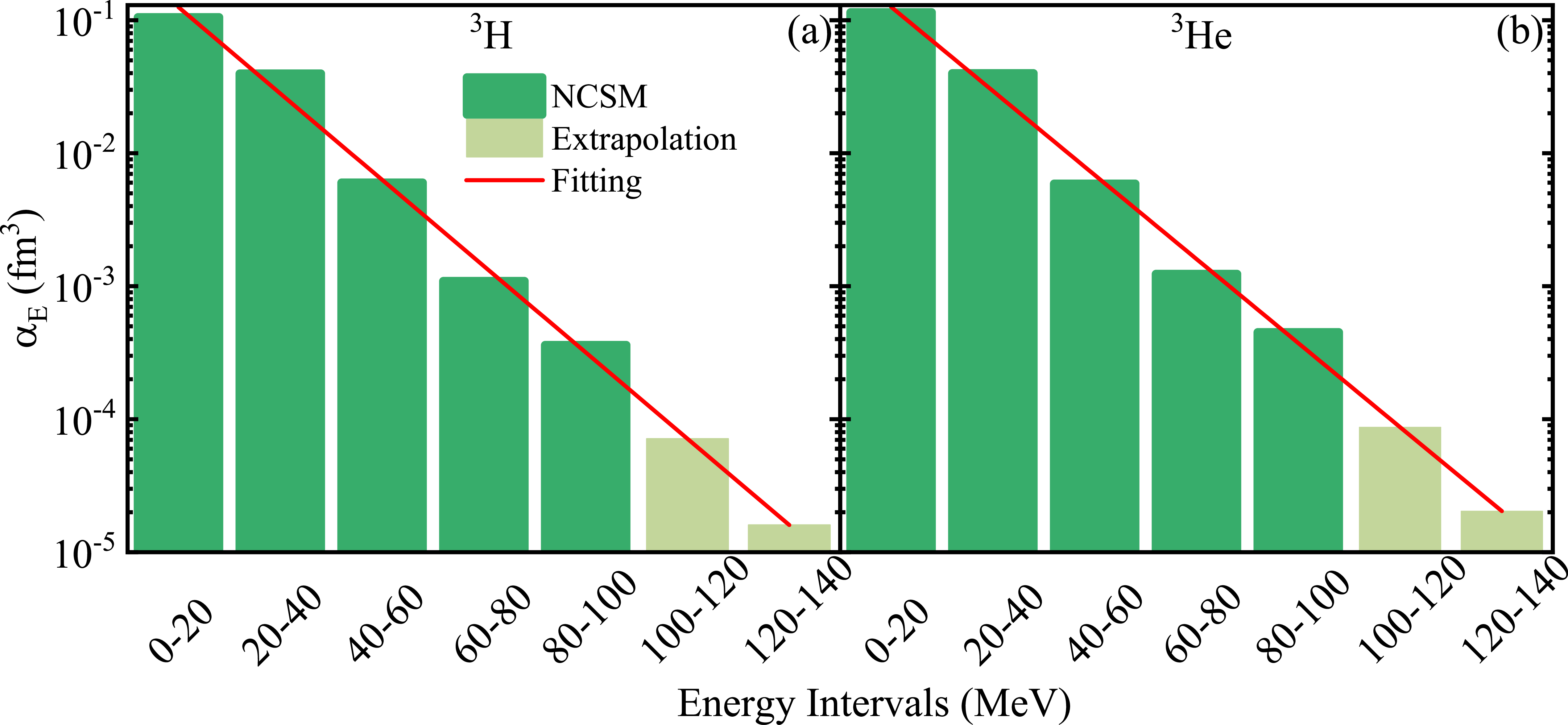}
 \caption{Contributions to $\alpha_E$ from successive excitation-energy intervals $E_a<E_x\le E_b$ at $N_{\max}=19$ and $\hbar\Omega=10$ MeV. Dark-green bins are NCSM results below 100 MeV; the red curve is obtained by exponential fitting to their amplitudes, and the two light-green bins are extrapolated from the fit.}
 \label{fig:bins}
\end{figure}

In order to estimate the uncertainty stemming from the excitation energy truncation, we fit the contributions to the polarizability $\alpha_E$ from successive excitation energy intervals of $20$ MeV by an exponential function, and extrapolate the fitting results up to the two intervals $100-120$ and $120-140$ MeV. The dark-green bins in Fig.~\ref{fig:bins} represent, as an example, the NCSM contributions obtained below $100$ MeV with $N_{\max}=19$ and $\hbar\Omega=10$ MeV, and the red curve is obtained by exponential fitting to their amplitudes. The light-green bins show the high-energy contribution obtained by extrapolating this fit beyond 100 MeV; the resulting total contribution above 100 MeV is about $0.0001$ fm$^3$ for both nuclei, accounting for only $\sim0.06\%$ of the total polarizability, which is included in the uncertainty analysis below.

\begin{figure}[!tbp]
 \centering
 \includegraphics[width=1.0\columnwidth]{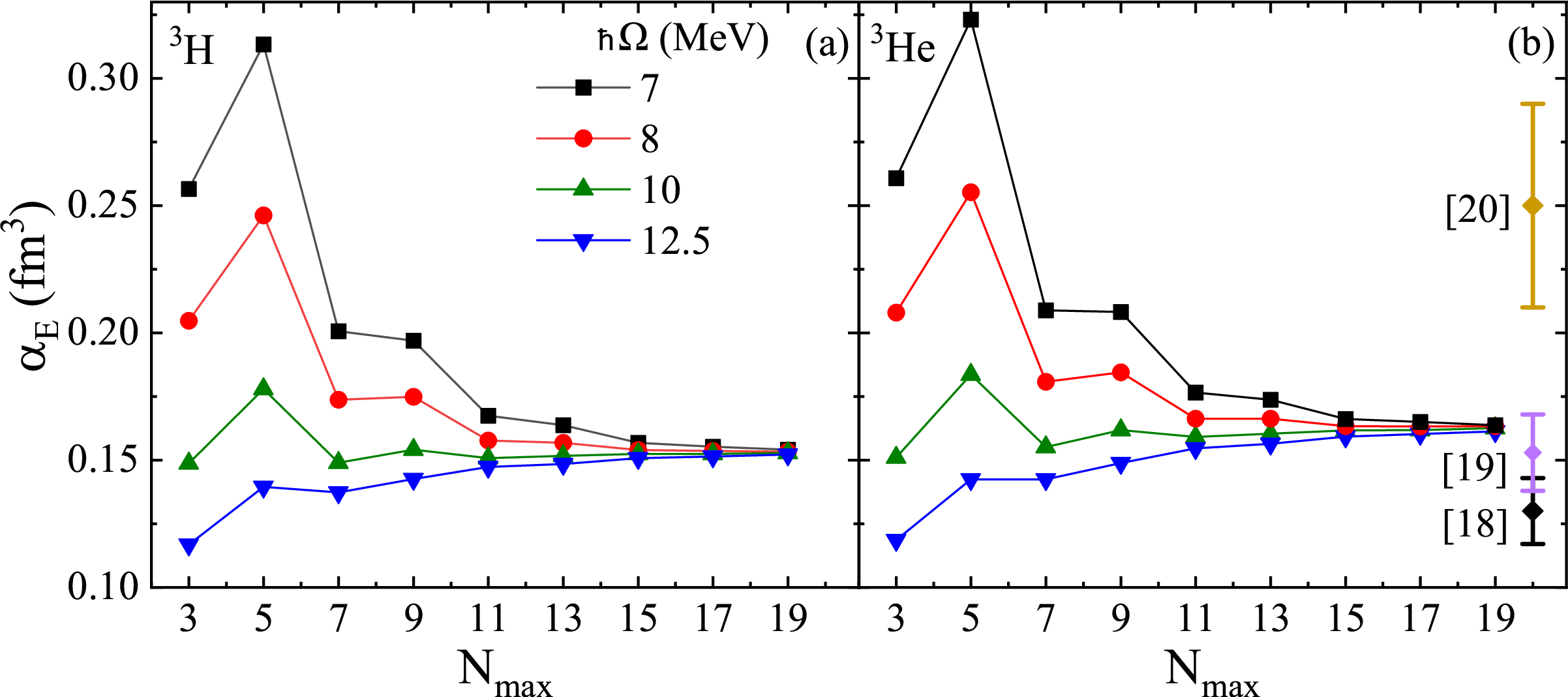}
 \caption{$E1$ polarizabilities of (a) $^3$H and (b) $^3$He as functions of the final-state $N_{\max}$ at the indicated $\hbar\Omega$ values.  The three experimental results for $^3$He, together with their quoted uncertainties, are taken from Refs.~\cite{Rinker1976,PachuckiMoro2007,Goeckner1991}.}
 \label{fig:convergence}
\end{figure}

We examine the model-space convergence of the total $E1$ polarizabilities of $^3$H and $^3$He in Fig.~\ref{fig:convergence} for several $\hbar\Omega$ values.  The total includes the $J_f^\pi=1/2^-$ and $3/2^-$ contributions, with the excitation spectrum truncated at $E_x^{\rm cut}=100$ MeV.  The pronounced $\hbar\Omega$ dependence at small $N_{\max}$ decreases as the model space grows, and the different $\hbar\Omega$ sequences approach common values.  For $^3$He, the largest-space results are compared with the three experimental determinations in panel (b), along with their quoted uncertainties.  This convergence demonstrates that the direct-summation approach is numerically feasible for the $E1$ response distributed over multiple final-state angular momenta.

Among these sequences, the $\hbar\Omega=10$ MeV results exhibit the most rapid convergence.  We therefore take the results at $N_{\max}=19$ and $\hbar\Omega=10$ MeV as our predictions for the polarizabilities accumulated up to $E_x^{\rm cut}=100$ MeV.  Following Ref.~\cite{Yin2024}, we estimate the numerical uncertainty from the excitation-energy truncation and the limited HO basis.  Extrapolating the fitted high-energy behavior in Fig.~\ref{fig:bins} gives contributions above $E_x^{\rm cut}=100$ MeV of approximately $0.0001$ fm$^3$ for both nuclei as discussed above.  To estimate the basis-truncation uncertainty, at $N_{\max}=19$ we take the difference between the maximum and minimum results over $\hbar\Omega=8$, 10, and 12.5 MeV.  These spreads are approximately $0.0012$ fm$^3$ for $^3$H and $0.0020$ fm$^3$ for $^3$He.  Since omitted high-energy strength can only increase $\alpha_E$, this contribution enters only the upper error bar: the upper uncertainty is the sum of the two components, whereas the lower uncertainty contains only the basis-truncation component.  The resulting $E1$ polarizabilities of $^3$H and $^3$He, including their estimated uncertainties, are summarized in Table~I.
The corresponding mirror difference is
$\Delta_C\alpha_E\equiv\alpha_E(^3{\rm He})-\alpha_E(^3{\rm H})=0.0097(8)$ fm$^3$, corresponding to an enhancement of approximately $6.3\%$ relative to $^3$H.  Its uncertainty is estimated primarily from the residual model-space dependence of the directly calculated mirror difference.  The excitation-energy-truncation uncertainty is negligible because the nearly equal extrapolated contributions above 100 MeV cancel at the quoted precision in the mirror difference.  For comparison, the $^3$H and $^3$He NCSM results of Ref.~\cite{Stetcu2009}, obtained with chiral $NN$ and three-nucleon ($3N$) interactions at N$^3$LO and N$^2$LO, respectively, imply $\Delta_C\alpha_E=0.010(5)$ fm$^3$ when their quoted values are subtracted and the uncertainties are combined in quadrature, consistent with the present result.  The Daejeon16 interaction is isospin symmetric, and the Coulomb interaction is the only isospin-breaking term retained in the present Hamiltonian.  Hence our calculated mirror difference is entirely Coulomb induced.
\begin{table}[!tbp]
\caption{$E1$ polarizabilities of the $A=3$ nuclei.  The theoretical approach and nuclear interaction are specified for each calculated result.  In the table, hyperspherical harmonics (HH), effective-interaction hyperspherical harmonics (EIHH), three-nucleon interaction ($3N$), Argonne v18 (AV18), Urbana IX (UIX), Trento potential (TN), and Malfliet--Tjon I/III (MT-I/III) are abbreviated as indicated.}
\label{tab:alpha}
\scriptsize
\setlength{\tabcolsep}{1.6pt}
\renewcommand{\arraystretch}{0.82}
\resizebox{\columnwidth}{!}{%
\begin{tabular}{@{}l@{\hspace{5pt}}llc@{}}
\hline\hline
Nucleus & Method & Interaction & $\alpha_E$ (fm$^3$) \\
\hline
$^3$H  & NCSM [This work] & $NN$ (Daejeon16) \cite{Shirokov2016} & $0.1529^{+0.0013}_{-0.0012}$ \\
       & NCSM \cite{Stetcu2009} & $NN$ (N$^3$LO) \cite{EntemMachleidt2003} + $3N$ (N$^2$LO) \cite{Navratil2007} & $0.139(2)$ \\
       & EIHH \cite{NevoDinur2016} & $NN$ (AV18) \cite{Wiringa1995} + $3N$ (UIX) \cite{Pudliner1997} & $0.137$ \\
       & EIHH \cite{NevoDinur2016} & $NN$ (N$^3$LO) \cite{EntemMachleidt2003} + $3N$ (N$^2$LO) \cite{Navratil2007} & $0.139$ \\
       & HH \cite{Efros1997} & $NN$ (TN) \cite{EfrosPRL1997} & $0.135$ \\
       & HH \cite{Efros1997} & $NN$ (MT-I/III) \cite{MalflietTjon1969} & $0.143$ \\
\hline
$^3$He & NCSM [This work] & $NN$ (Daejeon16) \cite{Shirokov2016} & $0.1626^{+0.0021}_{-0.0020}$ \\
       & NCSM \cite{Stetcu2009} & $NN$ (N$^3$LO) \cite{EntemMachleidt2003} + $3N$ (N$^2$LO) \cite{Navratil2007} & $0.149(5)$ \\
       & EIHH \cite{NevoDinur2016} & $NN$ (AV18) \cite{Wiringa1995} + $3N$ (UIX) \cite{Pudliner1997} & $0.149$ \\
       & EIHH \cite{NevoDinur2016} & $NN$ (N$^3$LO) \cite{EntemMachleidt2003} + $3N$ (N$^2$LO) \cite{Navratil2007} & $0.153$ \\
       & HH \cite{Efros1997} & $NN$ (TN)\cite{EfrosPRL1997} & $0.143$ \\
       & HH \cite{Efros1997} & $NN$ (MT-I/III) \cite{MalflietTjon1969} & $0.151$ \\
       & HH \cite{Leidemann2003} & $NN$ (AV18) \cite{Wiringa1995} + $3N$ (UIX) \cite{Pudliner1997} & $0.145$ \\
       & Experiment \cite{Rinker1976} & -- & $0.130(13)$ \\
       & Experiment \cite{PachuckiMoro2007} & -- & $0.153(15)$ \\
       & Experiment \cite{Goeckner1991} & -- & $0.250(40)$ \\
\hline\hline
\end{tabular}
}
\end{table}

We summarize our results in Table~\ref{tab:alpha} and compare them with previous calculations and experimental determinations. The quoted calculations employ different many-body methods and nuclear interactions, as specified in the table.  The previous theoretical values span 0.135--0.143 fm$^3$ for $^3$H and 0.143--0.153 fm$^3$ for $^3$He. The NCSM results of Ref.~\cite{Stetcu2009} were obtained with the Lanczos-moment method employing chiral $NN$ and $3N$ interactions. Our results are about 10\% higher than those of Ref.~\cite{Stetcu2009}, which suggests an overall systematic uncertainty in the theory arising from the dependence on the choice of Hamiltonian. Our results are also about 9\% higher than other previous calculations, but their consistency with the present results cannot be assessed due to the lack of quoted uncertainties. The three experimental $^3$He determinations, also included in Table~\ref{tab:alpha} with their quoted uncertainties, show some inconsistency, underscoring the need for improved constraints on the low-energy $E1$ response.

\begin{figure}[!tbp]
 \centering
 \includegraphics[width=1.0\columnwidth]{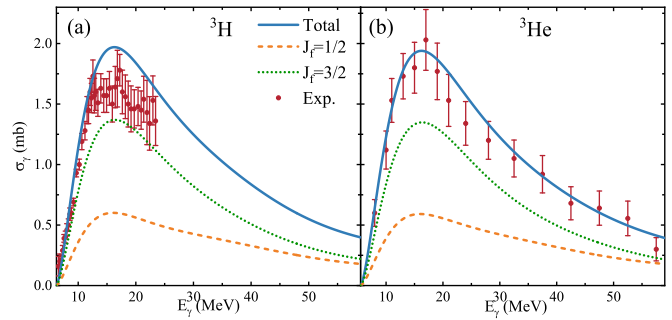}
 \caption{Photoabsorption cross sections of (a) $^3$H and (b) $^3$He reconstructed from the $N_{\max}=19$, $\hbar\Omega=10$ MeV finite spectra.  Total, $J_f^\pi=1/2^-$ and $3/2^-$ contributions are shown.  For each nucleus, the total experimental cross section (labeled Exp.) was constructed at photon energies common to its two measured breakup channels.  The $^3$H points combine the $^3$H$(\gamma,n)d$ and $^3$H$(\gamma,2n)p$ measurements of Ref.~\cite{Faul1981}; the $^3$He points combine the $^3$He$(\gamma,p)d$ and $^3$He$(\gamma,n)pp$ measurements of Ref.~\cite{Gorbunov1964}.}
 \label{fig:photo}
\end{figure}

We compare the photoabsorption cross sections reconstructed via the finite-spectrum LIT method~\cite{Yin2024} with the available data in Fig.~\ref{fig:photo}.  Both rise rapidly above the breakup threshold, exhibit a broad maximum near 16 MeV, and decrease smoothly at higher photon energy.  Around the maximum, the $3/2^-$ contribution is approximately twice the $1/2^-$ contribution, consistent with the angular-momentum hierarchy of the inverse-energy-weighted sums.  The calculated $^3$H cross section is systematically higher than much of the measured strengths near and above the maximum, whereas the $^3$He result is compatible with most of the data within their relatively large uncertainties.  As an internal consistency check, integrating the reconstructed cross sections with the nverse-energy-squared weight up to $E_\gamma=100$ MeV yields polarizabilities that differ from the corresponding direct-summation results by only $0.20\%$ for $^3$H and $0.29\%$ for $^3$He.  This sub-percent agreement shows that the LIT inversion preserves the inverse-energy-weighted strength to high accuracy.

The overall similarity of the mirror cross sections agrees with previous studies of three-nucleon photodisintegration \cite{Golak2000,Skibinski2003,Skibinski2005,Deltuva2004,Efros1997}.  The $^3$He response begins at a lower photon energy because Coulomb shifts the breakup threshold.  This displacement moves strength toward smaller excitation energies, where it is enhanced by the inverse-energy weighting in Eq.~(\ref{eq:alpha}), even though the two energy-dependent profiles remain similar.  The calculated curves contain the long-wavelength $E1$ contribution only, whereas the measurements represent total photoabsorption; retardation, higher multipoles, and current corrections may become relevant with increasing energy \cite{Leidemann2003}.

The inversion of the LIT was tested by varying the basis parameters, Lorentz width, excitation-energy cutoff, and the transform and response grids, following the analysis used for $^4$He \cite{Yin2026}.  The LIT inversion parameters used in calculations presented in Fig.~\ref{fig:photo} are discussed in Appendix~\ref{app:LIT}.

\section{Conclusions}

In summary, we have extended the \textit{ab initio} NCSM approach for calculating $E1$ polarizabilities $\alpha_E$ and photoabsorption cross sections, previously
applied to $^4$He to the nonzero-spin $A=3$ mirror nuclei. Using the Daejeon16 NN interaction, we calculate the $J_f^\pi=1/2^-$ and $3/2^-$ spectra and find that both contributions to the inverse-energy-weighted $E1$ response stabilize with increasing model-space size.  Including the finite-basis and excitation-energy-truncation uncertainties, we obtain $\alpha_E(^3{\rm H})=0.1529^{+0.0013}_{-0.0012}$ fm$^3$ and $\alpha_E(^3{\rm He})=0.1626^{+0.0021}_{-0.0020}$ fm$^3$.  The contribution of $J_f^\pi=3/2^-$ states to the polarizability is 2.12 times larger than that of the $J_f^\pi=1/2^-$ states in both nuclei.

The resulting mirror difference, $\Delta_C\alpha_E=0.0097(8)$ fm$^3$, corresponds to a $6.3\%$ enhancement in $^3$He relative to $^3$H.  Since the Daejeon16 NN interaction is isospin symmetric, the retained Coulomb term is the sole source of this difference.  The NCSM spectral information is used to construct and invert the finite-spectrum LIT and thereby obtain the photoabsorption cross sections.  These cross sections remain stable under the tested inversion variations and exhibit the same angular-momentum hierarchy as the polarizabilities.  Taken together, the polarizability and photoabsorption results establish the applicability of the finite-spectrum approach to integrated and energy-resolved $E1$ observables in nonzero-spin nuclei with multiple allowed final-state angular momenta.

\section*{Acknowledgments}

We acknowledge helpful discussions with Pieter Maris and Mark A. Caprio.  This work is partially supported by National Natural Science Foundation of China (No.~12675152), by Natural Science Foundation of Henan under Grant No.~252300421486, by the State Key Laboratory of Heavy Ion Science and Technology, Institute of Modern Physics, Chinese Academy of Sciences (Grant No.~HIST2026CO09), and National Key R\&D Program of China (2023YFA1606701).  A portion of the computational resources are provided by the National Energy Research Scientific Computing Center (NERSC), a U.S. Department of Energy Office of Science User Facility located at Lawrence Berkeley National Laboratory, operated under Contract No.~DE-AC02-05CH11231 using NERSC awards NP-ERCAP0028672, NP-ERCAP0032689 and NP-ERCAP0037058.  This work is partially supported by new faculty startup funding by the Institute of Modern Physics, Chinese Academy of Sciences, by Key Research Program of Frontier Sciences, Chinese Academy of Sciences, Grant No.~ZDB-SLY-7020, by the Natural Science Foundation of Gansu Province, China, Grant No.~20JR10RA067, by the Foundation for Key Talents of Gansu Province, by the Central Funds Guiding the Local Science and Technology Development of Gansu Province, Grant No.~22ZY1QA006, by international partnership program of the Chinese Academy of Sciences, Grant No.~016GJHZ2022103FN, by National Natural Science Foundation of China, Grant No.~12375143, by National Key R\&D Program of China, Grant No.~2023YFA1606903 and by the Strategic Priority Research Program of the Chinese Academy of Sciences, Grant No.~XDB34000000.  P. Yin and B. Zhou are supported by Shanghai Research Center for Theoretical Nuclear Physics, NSFC and Fudan University, Shanghai 200438, China and the National Natural Science Foundation of China under Grant No.~12147101.  A. M. Shirokov is thankful to the Chinese Academy of Sciences President's International Fellowship Initiative Program (Grant No.~2026PVA0079) which supported his visits to Lanzhou where a part of this work was performed and acknowledges the hospitality of Chinese colleagues during these visits.  A portion of the computational resources were also provided by Gansu Computing Center and Gansu Advanced Computing Center.  This material is based upon work supported by the U.S. Department of Energy, Office of Science, under Award No.~DE-FG02-95ER40934.

\appendix
\section{LIT inversion basis and parameters}
\label{app:LIT}

Following Ref.~\cite{Yin2026}, the reconstructed response is expanded as
\begin{equation}
 R(E_\gamma)\approx\sum_{n=1}^{N_b}c_n
 \chi_n(E_\gamma-E_{\rm th}),
 \label{eq:inversion_basis}
\end{equation}
with the basis functions
\begin{equation}
 \chi_n(E)=
 \begin{cases}
 E^{3/2}\exp[-E/(n\beta)], & E\ge 0,\\
 0, & E<0.
 \end{cases}
 \label{eq:basis_function}
\end{equation}
Here $E_{\rm th}$ is the calculated $d+n$ or $d+p$ separation threshold, $N_b$ is the number of basis functions, $\beta$ sets their energy scale, and the coefficients $c_n$ are determined by fitting the transform of Eq.~(\ref{eq:inversion_basis}) to the finite-spectrum LIT in Eq.~(\ref{eq:LIT}).  For the $N_{\max}=19$, $\hbar\Omega=10$ MeV spectra used in Fig.~\ref{fig:photo}, we employ six basis functions ($N_b=6$), a Lorentz width $\sigma_I=20$ MeV, and an excitation-energy cutoff $E_x^{\rm cut}=100$ MeV.  The basis scale is chosen as $\beta=3.75$ MeV for $^3$H and $\beta=4.00$ MeV for $^3$He.  The adopted values of $N_b$, $\sigma_I$, and $\beta$ were selected from stable regions identified in scans of the inversion parameters, following the procedure of Ref.~\cite{Yin2026}; for each nucleus, $\beta$ was chosen near the center of the corresponding stability region.

For a response reconstructed over $E_{\rm th}\le E_\gamma\le E_x^{\rm cut}$, the transform is sampled approximately over $E_{\rm th}-\sigma_I\le\sigma_R\le E_x^{\rm cut}+\sigma_I$, following Ref.~\cite{Andreasi2005}.  The resulting cross sections were verified to be stable against variations of the inversion parameters, enlargement of the $\sigma_R$ interval, and refinement of the numerical grids.

\end{document}